\documentclass[reprint,amsmath,amssymb,pra]{revtex4-1}
\usepackage{graphicx}
\usepackage{dcolumn}
\usepackage{bm}
\usepackage[usenames,dvipsnames]{xcolor} 
\usepackage[utf8]{inputenc}
\usepackage{latexsym}
\usepackage{amsfonts}
\usepackage{amssymb}
\usepackage{amsmath}
\usepackage{psfrag}
\usepackage{rotating}
\usepackage{pgf,pgfsys,pgffor} %
\usepackage{pgfplots}
\usepackage{pgfplotstable}
\usepackage{tikz}
\usetikzlibrary{intersections,arrows,decorations.pathmorphing,backgrounds,fit,mindmap,calc,through,scopes,fadings,positioning,automata,calendar,shapes,er,matrix,folding,patterns,petri,plothandlers,plotmarks,shadows,topaths,through,trees,pgfplots.units,decorations, snakes }  
\usepgfplotslibrary{fillbetween}
\graphicspath{{./figureColor/}} 
\newcommand{\AxisRotator}[1][rotate=0]{%
    \tikz [x=0.25cm,y=0.60cm,line width=.2ex,-stealth,#1] \draw (0,0) arc (-150:150:1 and 1);%
}
\tikzset{pbs/.style={path picture={%
	\draw[black]
	 (path picture bounding box.north west) -- (path picture bounding box.south east);
	}}} 
\tikzset{antiPbs/.style={path picture={%
	\draw[black]
	 (path picture bounding box.north east) -- (path picture bounding box.south west);
	}}} 
\tikzset{mirror/.style={path picture={%
	\draw[black]
	 (path picture bounding box.north west) -- (path picture bounding box.south east);
	}}}
	
	\usepackage{blkarray}
	
\newcommand{\tr}[2][]{\textrm{Tr}_{#1} \left[ {#2} \right]}

\newcommand{\ket}[1]{\vert {#1} \rangle}
\newcommand{\pure}[1]{\vert {#1} \rangle \langle {#1} \vert}

\newcommand{\san}[3]{\langle {#1} \vert {#2} \vert {#3} \rangle}
\newcommand{\ee}{e} 

\newcommand{\ii}{{i\mkern1mu}}

\newcommand{\beq}{\begin{equation}}
\newcommand{\eeq}{\end{equation}}

\begin{document}

\title{Naimark extension for the single-photon canonical phase measurement}
\author{Nicola Dalla Pozza}
\affiliation{Quantum Driving and Bio-complexity, Dipartimento di Fisica e Astronomia, Universit\`a degli Studi di Firenze, I-50019 Sesto Fiorentino, Italy}\email{nicola.dallapozza@unifi.it}
\author{Matteo G. A. Paris}%
\affiliation{Quantum Technology Lab, Dipartimento di Fisica 'Aldo Pontremoli', Universit\`a degli Studi di Milano, I-20133 Milano, Italy}\email{matteo.paris@fisica.unimi.it}%
\date{\today}
\begin{abstract}
We address the implementation of the positive operator-valued measure 
(POVM) describing the optimal $M$-outcomes discrimination of the 
polarization state of a single photon.  Initially, the POVM elements 
are extended to projective operators by Naimark theorem, then the 
resulting projective measure is 
implemented by a Knill-Laflamme-Milburn scheme involving an optical 
network and photon counters. 
We find the analytical expression of the Naimark extension and the detection scheme that realise it for an arbitrary number of outcomes $M=2^N$.
\end{abstract}
\maketitle
\section{\label{introduction}Introduction}
Quantum information processing with flying qubits, especially photons,  
has been extensively studied in the last decades, with applications in
communication \cite{Gisin2007, Gyongyosi2018}, quantum key distribution 
(QKD) \cite{Gisin2002, Lo2014}, quantum networking \cite{Elliott2002, Lloyd2004, Kimble2008} and universal quantum computing \cite{Knill2001, Raussendorf2001, Ralph2006}. While remarkable results for communication and QKD have already been proved \cite{Chen2012, Pfister2018, Boaron2018, Zhang2018}, there is great expectation 
for the medium-to-long-term realization of a quantum internet \cite{Dahlberg2018, Wehner2018}, and possibly of a photonic quantum computer \cite{Takeda2019}.

In all these scenarios, photons are used as qubit encoders and information 
carriers because of their ability to travel long distances with minimal 
decoherence. Information processing is performed at the encoding and 
decoding stage, and in intermediate stages as well (e.g. in repeaters 
\cite{Briegel1998, Azuma2015}) to implement quantum gates. Depending on the 
qubit encoding into the possible degrees of freedom, gates may have easy
or more challenging implementations in terms of resources \cite{Ralph2005}, 
with two-qubit gates being the hardest components to realize due to the 
small photon-photon coupling that can be obtained via matter-mediated 
processes \cite{Thompson2011}.

Several theoretical and experimental proposals have been put forward 
\cite{OBrien2009, Northup2014, Diamanti2016, ShenoyHejamadi2017, Pirandola2019, 
Kok2007, Krovi2017}, but the transition from theory to practice is not always straightforward. 
In this paper, we focus our attention on quantum measurements 
and consider a detection scheme which arises in optimal discrimination theory.
The detection operators are described by an analytical expression, 
and we translate it into an optical scheme. We believe the steps undertaken in this paper 
are relevant in the framework of current technology and
that they will prove useful for the implementation of other measurement schemes 
in linear optics quantum computing with discrete variables.
\par
The most general description of a quantum measurement 
is provided by positive operator-valued measures (POVMs) 
acting on the Hilbert space of the systems under 
investigation. Mathematically speaking, a POVM is a 
resolution of identity made by sets of positive operators 
$\{\Pi_k\}$, $\sum_k \Pi_k = {\mathbb I}$. The cardinality 
of the POVM is not limited by the dimension of the Hilbert
space and the operators $\{\Pi_k\}$ are not required to
be projectors. POVMs have found useful applications in several 
fields of quantum information theory, e.g. in unambiguous 
quantum discrimination, where the optimal detection scheme 
may not correspond to a projective measurement.
\par
Experimental realizations, however, always involve {\em observable}
quantities, which strictly correspond to projective measures. Therefore,
the challenge usually comes in finding an appropriate detection scheme to realize
 a given POVM. Fortunately, there is a canonical route 
to achieve this goal, which is provided by the Naimark theorem \cite{Naimark1940, Akhiezer1993, Helstrom1973, Helstrom1976, Holevo2001}, ensuring 
that the POVM formulation may always be extended to a projecive one, which
may then be implemented experimentally. 
More explicitly, the Naimark
theorem states that for any POVM $\{\Pi_k\}_{k\in{\cal K}}$ on the 
Hilbert space $\mathcal{H}_S$, generating a probability 
distribution $p(k) = \hbox{Tr}_{S}[\rho\, \Pi_k]$, 
$\forall \rho$, there 
exist a set of orthogonal projectors $\{P_k\}_{k\in{\cal K}}$ on
the enlarged Hilbert space $\mathcal{H}_A \otimes \mathcal{H}_S$ and a pure
state $|\omega\rangle\in \mathcal{H}_A$ such that 
$$
p(k)=\hbox{Tr}_{S}[\rho\, \Pi_k]
=\hbox{Tr}_{AS}[\rho\otimes |\omega\rangle\langle\omega|\, P_k]\ .
$$

\par
In this paper, we address the implementation of the POVM $\{\Pi_k\}$, 
$k= 0,1,\ldots ,M-1$ which acts on the Hilbert space 
of a two-level system, and describes the optimal $M$-outcomes 
discrimination of the polarization state of a single photon. 
The explicit expression is given by 
\begin{align}
\Pi_k & = \frac{2}{M}\,
\pure{\psi_k} \label{povm}\\ 
|\psi_k\rangle & = \frac1{\sqrt{2}} 
\left(\ee^{-i \frac\pi{M} k} |0\rangle + \ee^{i \frac{\pi}{M}k}|1\rangle\right)\,.
\end{align} 
Our strategy to solve the problem is the following: firstly, the 
POVM elements $\Pi_k$ are extended to projective operators upon
exploting the Naimark theorem; secondly the resulting projective measure 
is implemented by a Knill-Laflamme-Milburn scheme involving an optical 
network and photon counters. The key idea is to consider an interferometer 
with a single rail input mode to receive the input signal, and multiple 
output path modes, one for each outcome. Each of these output modes goes to 
photon counters, such that when we record a click we assign the 
corresponding outcome. In this way, we have found explicitly the 
Naimark extension and the detection scheme for an arbitrary number of outcomes
$M=2^N$, $N>1$. 
\par
The paper is structured as follows. In Section \ref{sec:Naimark} 
the Naimark theorem is introduced and the algorithm to evaluate the extension of the phase measurement is described.
In Section \ref{application} the application of the Naimark extension to the phase measurement of the polarization of a single photon is presented. In Section \ref{Zconstruction} we give the corresponding expressions in the case of $M=8$.
The extension is factorized in a sequence of unitaries in Section \ref{decomposition}, with Section \ref{Zdecomposition} considering the case of $M=8$ in this instance. The implementation of the unitaries are provided in Section \ref{physicalRealization}, and two different schemes for the phase measurement are designed in Sections \ref{direct} and \ref{folded}. Section IV closes the paper with some concluding remarks.
\section{Naimark extension of phase POVM}
\label{sec:Naimark}
We are going to consider phase-measurements on a qubit. 
In particular, we consider measurements described by 
the POVM $\{\Pi_k\}$, where the $k$-th outcome,  $k \in \{0, 1, 
\ldots, M-1\}$,  is associated with the phase 
value $\theta_k = k\frac{2\pi}{M}$. The number of outcomes 
$M$ defines the resolution of the measurement scheme, and it 
may be arbitrarily high. The scheme we are going to
discuss works for $M$ being a power of 2, i.e. $M=2^N, N \geq 1 
\in \mathbb{N}$.
\par
As a matter of fact, many different kind of phase measurements 
have been analysed and discussed, with the main goal of achieving
optimal phase estimation. In this paper we study how to 
implement the phase measurement which is the solution 
of the following problem. Given the states $$\ket{\varphi_k} = \frac{\ket{0} + \ee^{i \varphi_k} \ket{1}}{\sqrt{2}}\,,$$ 
where $\varphi_k = \frac{2\pi}{M}k$, $ k \in \{0, 1, \ldots, M-1\}$, 
drawn with equal probability $\frac{1}{M}$, find the optimal 
POVM $\{\Pi_k\}$ that maximizes the probability of 
guessing correctly, i.e.
\beq
P_{guessing} = \sum_k P[\theta_k | \varphi_k] = \sum_k \san{\varphi_k}{\Pi_k}{\varphi_k} \ .
\label{correctProbability}
\eeq
The problem is well known because of the symmetry of the states, 
and the solution, which was found long ago \cite{Helstrom1976}, is 
the POVM in Eq. (\ref{povm}).  This optimal POVM may also be seen as 
an approximate \emph{canonical phase-measurement}, that is, the measurement 
defined by the POVM $\{\Pi_\theta\},\ \theta~\in~[0,~2\pi)$ defined in the standard basis as
\beq
\Pi_\theta = \frac{1}{2 \pi}\,\pure{\theta}\,,
\quad \ket{\theta} = \sum_n \ee^{\ii n \theta} \ket{n}\ . 
\label{canonicalPhaseMeasurement}
\eeq
When we restrict the Hilbert space to the subspace spanned by $\ket{n}=\{\ket{0},\ \ket{1}\}$ and we discretize the outcome $\theta$ in the $M$ values $\theta_k$, we obtain the POVM $\{\Pi_k\}$ of Eq. (\ref{povm}).
Note that each POVM element may be expressed in terms of the 
column vectors
\beq
\ket{\psi_k} = \frac{1}{\sqrt{2}}
\begin{bmatrix}
\ee^{-i k\frac{\pi}{M}} \\
\ee^{i k \frac{\pi}{M}} 
\end{bmatrix}
\label{phaseMeasurementPOVM}
\eeq
for $k= 0,1,\ldots ,M-1$, and also  as
\beq
\Pi_k = X_k X_k^{\dagger}, \quad
X_k=\frac{1}{\sqrt{M}} 
\begin{bmatrix}
\ee^{-i k\frac{\pi}{M}} \\
\ee^{i k \frac{\pi}{M}} 
\end{bmatrix}\,,
\eeq
i.e., using the set of the \emph{unnormalized} column vector $X_k$.  
Note also that the POVM elements are not orthogonal, except 
for $M=2$, $\Pi_k \Pi_l \neq \Pi_k \delta_{k,l}$.
\par
The POVM elements are operators on the original system Hilbert 
space $\mathcal{H}_S$ and, according to the Naimark Theorem \cite{Naimark1940, Akhiezer1993, Helstrom1973, Helstrom1976, Holevo2001}, may be implemented as a \emph{projective measurement} in a larger 
Hilbert space $\mathcal{H}$, usually referred to as the 
\emph{Naimark extension} of the POVM. Actually, the theorem ensures 
that a \emph{canonical extension} exists amongst the infinite others, 
i.e., an implementation as an indirect measurement, where the system 
under investigation is coupled to an independently prepared probe system \cite{Peres1990}
and then only the probe is subject to a (projective) measurement \cite{He2007, Bergou2010, Paris2012}, whose
statistics mimick that of the POVM.
\par
Here we look for a Naimark extension of the POVM 
$\{\Pi_k\}$ in Eq. (\ref{povm}) using a recursive algorithm designed 
in a previous paper \cite{DallaPozza2017}. The algorithm builds 
the projectors one by one, enlarging the size of the Hilbert space 
$\mathcal{H}$ only when necessary. As we will see in a moment, the 
projectors have rank one and may be described in a matrix representation 
as $P_k = Z_k Z_k^{\dagger}$, with $Z_k$ a column vector. 
Each projector must verify a set of orthogonality conditions, 
which translates to constraints on $Z_k$, i.e.
\beq
P_k P_l = 0, \ k\neq l \quad \iff \quad Z_k^\dagger Z_l = 0 \ ,
\label{orthogonalityCondition}
\eeq
as well as an idempotent condition,
\beq
(P_k )^2 = P_k \quad \iff \quad Z_k^\dagger Z_k = 1 \ .
\label{idempotentCondition}
\eeq
In addition, in order to be the extension of a POVM element, each projector $P_k$ must satisfy 
\beq
\Pi_k = \tr[A]{P_k ( \rho_A \otimes \mathbb{I}_S) },
\label{POVMrelation}
\eeq
which is the constraint required to evaluate the correct outcome probability in $\mathcal{H}_S$ and in $\mathcal{H}$ respectively, i.e.
\beq
\tr[S]{\Pi_k \rho_S} = \tr[AS]{P_k ( \rho_A \otimes \rho_S) }\ .
\label{sameOutcomeProbabilityConstraint}
\eeq
In Eqs. \eqref{POVMrelation} and in \eqref{sameOutcomeProbabilityConstraint}, we have introduced the enlarged Hilbert space $\mathcal{H}$ given by the tensor product of an ancillary Hilbert space $\mathcal{H}_A$ and the original Hilbert space $\mathcal{H}_S$, i.e. $\mathcal{H} = \mathcal{H}_A \otimes \mathcal{H}_S$. We have introduced also an auxiliary state $\rho_A$ defined in $\mathcal{H}_A $, 
whose choice gives some degrees of freedom in building the extension. 
Following the suggestion of Helstrom \cite{Helstrom1976}, we use $\rho_A = \pure{e_1^{\mathcal{A}}}$, where $\ket{e_1^{\mathcal{A}}}$ is the ancillary pure state whose column representation is the vector $e_1$ of the canonical basis with the appropriate size \footnote{Since the algorithm enlarges the size of the Hilbert space $\mathcal{H}$ only when necessary, the size of $\mathcal{H}_A$ is only determined at the end.}, i.e., with all the entries equal to zero except for the first one. 

The recursive algorithm works by building the columns $Z_k$ one at a time. In each column, the upper coefficients are set equal to $X_k$ \footnote{By Eq.~\eqref{POVMrelation}, the choice \unexpanded{$\rho_A = \pure{e_1^{\mathcal{A}}}$} imposes that the first entries of $Z_k$ are equal to $X_k$.}. Then, the following coefficients are found by imposing the orthogonality condition \eqref{orthogonalityCondition} with the previously found $Z_0, \ldots, Z_{k-1}$. Finally, the last coefficient is obtained by solving Eq.~\eqref{idempotentCondition}. All following coefficients are set to zero. 
If during the evaluation of a coefficient the provisional vector is already orthogonal to $Z_l,\ l<k$ or idempotent, it is not necessary to add another coefficient. This helps in reducing the growth in size of $\mathcal{H}$. The paper \cite{DallaPozza2017} actually finds a general   expression for the coefficients to solve the orthogonal and idempotent constraints.
The algorithm can be implemented numerically to find the Naimark extension of the POVM $\{\Pi_k\}$. However, an analytical expression for the projectors for arbitrarily high $M$ can be found when employing the order $Z_0$, $Z_{M/2}$, $Z_1$, $Z_{M/2+1}$, \dots, $Z_{M-1}$ for their evaluation.
The projectors are hence extended in pairs evaluating $Z_k$ and $Z_{k + M/2}$ for $k=0,\dots,M/2-1$, resulting in the overall expressions of Eq. \eqref{Zk}. 
To better illustrate how the recursive algorithm works, we show the case for $M=8$ in the Section \ref{Zconstruction}.

As expected, the first two coefficients are $X_k$. Then, $2(k+1)$ coefficients are defined, followed by zero entries that pad the vector up to the size of $M$. For $k=M/2-1$ and $k=M-1$ the last coefficients (which would overflow the length of $M$) are zeros, so that $Z_{M/2-1}$ and $Z_{M-1}$ can be truncated to the correct length.
The columns $Z_k$ can be packed in the $M \times M$ matrix $Z$, 
\beq
Z = \left[ Z_0 \ Z_{M/2} \ Z_1 \ Z_{M/2+1} \ \cdots \ Z_{M/2-1} \ Z_{M-1} \right]
\label{matrixZ}
\eeq
and Eqs.~\eqref{orthogonalityCondition}, \eqref{idempotentCondition} can be checked analitically or numerically to verify $Z^{\dagger} \cdot Z = Z \cdot Z^{\dagger} = I$. As an example, we evaluate $Z$ for $M=8$ in Section \ref{Zconstruction} and report it in Eq.~\eqref{matrixZ8}. Expression \eqref{Zk} may be proved by induction.
\begin{widetext}
\beq
Z_k =\begin{bmatrix}
\frac{\ee^{-i\frac{k}{M}\pi}}{\sqrt{M}} \\
\frac{\ee^{i\frac{k}{M}\pi}}{\sqrt{M}} \\
- \frac{2}{\sqrt{M(M-2)}}\cos \left(\frac{k}{M} \pi \right)\\
-\frac{2}{\sqrt{M(M-2)}}\sin \left(\frac{k}{M} \pi\right) \\
- \frac{2}{\sqrt{(M-2)(M-4)}}\cos \left(\frac{k-1}{M} \pi\right)\\
- \frac{2}{\sqrt{(M-2)(M-4)}}\sin \left(\frac{k-1}{M} \pi\right) \\
\vdots \\
- \frac{2}{\sqrt{(M-2k+2)(M-2k)}}\cos \left(\frac{1}{M} \pi\right)\\
- \frac{2}{\sqrt{(M-2k+2)(M-2k)}}\sin \left(\frac{1}{M} \pi\right) \\
\sqrt{\frac{M-2k-2)}{M-2k}} \\
0 \\
0\\
\vdots\\
0
\end{bmatrix}
, \quad
Z_{k+M/2} =\begin{bmatrix}
\frac{\ee^{-i\frac{k+M/2}{M}\pi}}{\sqrt{M}} \\
\frac{\ee^{i\frac{k+M/2}{M}\pi}}{\sqrt{M}} \\
\frac{2}{\sqrt{M(M-2)}}\sin \left(\frac{k}{M} \pi\right)\\
- \frac{2}{\sqrt{M(M-2)}}\cos \left(\frac{k}{M} \pi\right) \\
\frac{2}{\sqrt{(M-2)(M-4)}}\sin \left(\frac{k-1}{M} \pi\right)\\
- \frac{2}{\sqrt{(M-2)(M-4)}}\cos \left(\frac{k-1}{M} \pi\right) \\
\vdots \\
\frac{2}{\sqrt{(M-2k+2)(M-2k)}}\sin \left(\frac{1}{M} \pi\right)\\
- \frac{2}{\sqrt{(M-2k+2)(M-2k)}}\cos \left(\frac{1}{M} \pi\right) \\
0\\
\sqrt{\frac{M-2k-2}{M-2k}} \\
0\\
\vdots\\
0
\end{bmatrix}
\label{Zk}
\eeq
\end{widetext}

\subsection{\label{Zconstruction}Recursive evaluation of $Z_k$ for $M=8$}

In this section we illustrate how the recursive algorithm presented in \cite{DallaPozza2017} builds the columns of $Z$ for $M=8$. The procedure can be followed by looking at the columns of Eq. \eqref{matrixZ8}, which represent the resulting matrix.

As anticipated in Section \ref{sec:Naimark}, the columns are evaluated one at a time starting from $Z_0$ and following the order of Eq. \eqref{matrixZ}. In $Z_0$, the first two coefficients are $X_0 = [1/\sqrt{M},\ 1/\sqrt{M} ]^T$. Then, the next coefficient is obtained by imposing the condition that the overall $Z_0$ has unitary norm, as in Eq.~\eqref{idempotentCondition}. These three coefficients will be extended and padded with zeros once the final length is known. 

For the second column, which corresponds to $Z_4$, the first two coefficients are $X_4 = [-\ii/\sqrt{M},\  \ii/\sqrt{M}]^T$. The following one is obtained by imposing the orthogonality constraint \eqref{orthogonalityCondition} with $Z_0$, which gives a zero coefficient in the third item. The next coefficient is obtained from \eqref{idempotentCondition} imposing the unit norm.

The third column, which correspond to $Z_1$, is evaluated with the same procedure. The first two coefficients are $X_1=[\ee^{-i \pi /8 }/\sqrt{M},\ \ee^{i \pi /8 }/\sqrt{M}]^T$. The following coefficients are obtained from the orthogonality constraint with $Z_0$ and $Z_4$, obtaining $-2\cos \left(\pi /8 \right)/\sqrt{(M-2)M}$ and $-2 \sin \left(\pi /8 \right)/\sqrt{(M-2)M}$, respectively. The following coefficient is obtained again from the idempotent constraint \eqref{idempotentCondition}.

The recursive procedure continues in the same way for the remaining columns, first by copying the coefficients of $X_k$, then by imposing the orthogonal constraint  \eqref{orthogonalityCondition} with all the previous columns, and finally evaluating the last coefficient by the idempotent constraint \eqref{idempotentCondition}.

Note that while with this procedure up to $M+2$ coefficients may be evaluated for each column, the last coefficients of the last two columns are zeros since $M=8$, and the columns can be truncated to the correct length of $M=8$.

\begin{widetext}
\beq
Z =\begin{bmatrix}
\frac{1}{\sqrt{M}} & -\frac{i}{\sqrt{M}} & \frac{\ee^{-i \pi /8 }}{\sqrt{M}} & \frac{\ee^{-i5 \pi /8 }}{\sqrt{M}} & \frac{\ee^{-i 2 \pi /8 }}{\sqrt{M}} & \frac{\ee^{-i6 \pi /8 }}{\sqrt{M}} & \frac{\ee^{-i3 \pi /8 }}{\sqrt{M}} & \frac{\ee^{-i 7 \pi /8 }}{\sqrt{M}} \\

 \frac{1}{\sqrt{M}} &  \frac{i}{\sqrt{M}} & \frac{\ee^{i \pi /8 }}{\sqrt{M}} &  \frac{\ee^{i5 \pi /8 }}{\sqrt{M}} & \frac{\ee^{i2 \pi /8 }}{\sqrt{M}} & \frac{\ee^{i6 \pi /8 }}{\sqrt{M}} & \frac{\ee^{i3 \pi /8 }}{\sqrt{M}} & \frac{\ee^{i7 \pi /8 }}{\sqrt{M}} \\

 \sqrt{\frac{M-2}{M}} &  0 &  -\frac{2\cos \left(\pi /8 \right) }{\sqrt{(M-2)M}}   &  \frac{2 \sin \left(\pi /8 \right)}{\sqrt{(M-2)M}}  & - \frac{2 \cos \left(2\pi /8 \right)}{\sqrt{M(M-2)}} &  \frac{2 \sin \left(2\pi /8 \right)}{\sqrt{M(M-2)}}& - \frac{2 \cos \left(3\pi /8 \right)}{\sqrt{M(M-2)}}& \frac{2 \sin \left(3\pi /8 \right)}{\sqrt{M(M-2)}}\\

 0 & \sqrt{\frac{M-2}{M}} &  -\frac{2 \sin \left(\pi /8 \right)}{\sqrt{(M-2)M}}  &  -\frac{2  \cos \left(\pi /8 \right) }{\sqrt{(M-2)M}}& - \frac{2 \sin \left(2\pi /8 \right)}{\sqrt{M(M-2)}} & - \frac{2 \cos \left(2\pi /8 \right)}{\sqrt{M(M-2)}} & - \frac{2 \sin \left(3\pi /8 \right)}{\sqrt{M(M-2)}} & - \frac{2 \cos \left(3\pi /8 \right)}{\sqrt{M(M-2)}} \\

 0 & 0 &  \sqrt{\frac{M-4}{M-2}} &  0 & - \frac{2\cos \left(\pi /8 \right)}{\sqrt{(M-2)(M-4)}}& \frac{2 \sin \left(\pi /8 \right)}{\sqrt{(M-2)(M-4)}}& - \frac{2 \cos \left(2\pi /8 \right)}{\sqrt{(M-2)(M-4)}} & \frac{2\sin \left(2\pi /8 \right)}{\sqrt{(M-2)(M-4)}}\\

 0 &  0 & 0 &  \sqrt{\frac{M-4}{M-2}} & - \frac{2 \sin \left(\pi /8 \right)}{\sqrt{(M-2)(M-4)}} & - \frac{2 \cos \left(\pi /8 \right)}{\sqrt{(M-2)(M-4)}} & - \frac{2 \sin \left(2\pi /8 \right)}{\sqrt{(M-2)(M-4)}} & - \frac{2 \cos \left(2\pi /8 \right)}{\sqrt{(M-2)(M-4)}} \\ 

 0 &  0 & 0 & 0 & \sqrt{\frac{M-6}{M-4}} & 0& - \frac{2 \cos \left(\pi /8 \right)}{\sqrt{(M-4)(M-6)}}& \frac{2 \sin \left(\pi /8 \right)}{\sqrt{(M-4)(M-6)}}\\ 

 0 &  0 & 0 & 0 & 0 & \sqrt{\frac{M-6}{M-4}}& - \frac{2 \sin \left(\pi /8 \right)}{\sqrt{(M-4)(M-6)}} & - \frac{2 \cos \left(\pi /8 \right)}{\sqrt{(M-4)(M-6)}} \\

 0 &  0 & 0 & 0 & 0 & 0 & \sqrt{\frac{M-8}{M-6}} & 0 \\
 0 &  0 & 0 & 0 & 0 & 0 & 0 & \sqrt{\frac{M-8}{M-6}} 
\end{bmatrix}
\label{matrixZ8}
\eeq
\end{widetext}

\section{Phase-Measurement on a Single Photon}
\label{application}
Let us denote by $\rho_S$ the state of the qubit, defined on a two-level system 
representing the \emph{single rail polarization encoding}, that is, identifying the logical system-basis $\{\ket{0}_L,\ \ket{1}_L\}$ with the polarization modes $\ket{0}_L = \ket{H} = a^{\dagger (m)}_{H} \ket{0},\ \ket{1}_L = \ket{V} = a^{\dagger (m)}_{V} \ket{0}$. Operators $a^{\dagger (m)}_{H},\ a^{\dagger (m)}_{V}$ are the creation operators of the polarization modes on the $m$-th path and $\ket{0}$ is the vacuum state. In this case the optical state is defined on a \emph{single} path, as opposed to the \emph{dual rail} encoding which employs the $m$-th and $n$-th spatial modes to define the logical basis $\ket{0}_L = \ket{10}_{mn} = a^{\dagger(m)} \ket{00}_{mn}$ and $ \ket{1}_L = \ket{01}_{mn} = a^{\dagger(n)} \ket{00}_{mn}$. 
\par
Note that even though our system qubit is defined with the single rail 
polarization encoding, in the following we will also employ 
the dual rail encoding when speaking about the implementation scheme 
of the phase measurement. In that framework, we will denote the mode number in the superscript, while making  the polarization explicit in the subscript.
\par
Let us start by summarising the key idea behind our detection scheme. 
We implement the Naimark extension of the POVM by an optical network that 
receives the quantum state $\rho_S$ to be measured as input, and 
(probabilistically) outputs a single photon towards an array of photon 
counters. Each detector is associated with an outcome, corresponding to a 
click in a specific detector. In the ideal case of no losses in the network,
and no detector noise, every time we send $\rho_S$ into the optical network 
we always get one and only one click. 
In this respect, our scheme resembles the KLM scheme for measurements
\cite{Knill2001} since the measurement device is implemented 
with a unitary rotation followed by a set of projectors.
In our case the overall projectors $P_k$ to be applied on $\rho_A \otimes \rho_S$ 
can be obtained as $P_k = Z \pure{e_k^{\mathcal{H}}} Z^{\dagger}$, where $\ket{e_k^{\mathcal{H}}}$ is the state defined in $\mathcal{H}$ with column representation of the $k$-th element of the canonical basis. Since
\beq 
\tr[]{P_k \ \rho_A \otimes \rho_S} = \tr[]{ \left( Z^{\dagger} \rho_A \otimes \rho_S Z \right)\pure{e_k^{\mathcal{H}}}},
\eeq
the unitary rotation is defined by $Z^{\dagger}$ and implemented with the optical network, while the projector $\pure{e_k^{\mathcal{H}}}$ is implemented with a photon counter on the $k$-th output mode.
\subsection{Decomposition of the unitary $Z^{\dagger}$}
\label{decomposition}
The unitary $Z^{\dagger}$ can be decomposed as a product of simpler unitary 
rotations \cite{HornJohnson} usually referred to as \emph{Givens Rotations} 
(GR), i.e., a rotation in the plane spanned by two coordinates axes, often employed 
to zero out a particular entry in a vector. 
Section \ref{physicalRealization} describes how to implement each GR, so that the overall sequence realizes the interferometer associated with $Z^{\dagger}$.
GR have a matrix representation that looks like the identity matrix, with the exception of the coefficients on two rows and two columns, which define the mixing between the two. 

We define such an $M \times M$ matrix as
\beq
W(u,v, \omega) = \ \begin{blockarray}{cccccc}
 &  & u & v &  \\
\begin{block}{(ccccc)c}
  1 & 0 & 0 & 0 & 0 &  \\
  0 & \ddots & 0 & 0 & 0 &  \\
  0 & 0 & \cos(\omega) & \sin(\omega) & 0 & \ u \\
  0 & 0 & -\sin(\omega) & \cos(\omega) & 0 & \ v \\
  0 & 0 & 0 & 0 & 1 &  \\
\end{block}
\end{blockarray}
\eeq
with $u,v$ the indices of the rows and columns being mixed, and $\omega$ 
a parameter defining the mixing. We define also the matrix $S(u, \phi)$
\beq
S(u, \phi) = \ \begin{blockarray}{cccccc}
 &  &  & u &  \\
\begin{block}{(ccccc)c}
  1 & 0 & 0 & 0 & 0 &  \\
  0 & \ddots &  & \vdots &  &  \\
  0 & & 1 & 0 & 0 &  \\
  0 & \cdots & 0 & \ee^{-i\phi} & 0 & \ u \\
  0 &  & 0 & 0 & 1 &  \\
\end{block}
\end{blockarray}\,,
\eeq
which corresponds to a phase shift on the $u$-th vector of the basis. 
\par

In decomposing $Z^{\dagger}$ we take advantage of its structure, which is almost lower-triangular due to the zero padding of $Z_k$ to reach the length of $M$ coefficients (see for instance the structure of $Z$  in Eq.~\eqref{matrixZ8} in the example in Section~\ref{Zconstruction}). For further details on the decomposition of $Z^{\dagger}$ are reported in Section~\ref{Zdecomposition}.

A pattern in the sequence of unitaries $W$ and $S$ emerges, suggesting an analytical expression for the decomposition for any $M$ (see for instance Eq.~\eqref{decomposition8}). In fact, with the exception of $S(2, \pi/2)$ and $W(1,2,\pi/4)$, the GR can be grouped in triplets of unitaries where the $u,\ v$ indices act on the same group, e.g. \{5,7,6,8\}, \{3,4,5,6\} or \{1,2,3,4\}. The parameter $\omega$ also shows a pattern in its value, i.e. it has the same value in the first two GR of the triplet and it has the same value in the third GR amongst different triplets. These patterns have a direct effect on the physical realization of $Z^{\dagger}$ decomposition (see Section \ref{physicalRealization}).

\par
\begin{widetext}
\begin{align}
Z^{\dagger} = & W\left(7,8,\pi + \frac{\pi}{M} \right) \cdot 
W\left(6,8,\arctan\sqrt{\frac{M-6}{2}}\right) \cdot 
W\left(5,7,\arctan\sqrt{\frac{M-6}{2}}\right) \nonumber \\
	& \cdot W\left(5,6,\pi + \frac{\pi }{M} \right) \cdot W\left(4,6,
   \arctan\sqrt{\frac{M-4}{2}}\right) \cdot W\left(3,5,\arctan
   \sqrt{\frac{M-4}{2}}\right) \nonumber \\
	& \cdot W\left(3,4,\pi + \frac{\pi }{M} \right) \cdot W\left(2,4,
   \arctan \sqrt{\frac{M-2}{2}}\right) \cdot W\left(1,3,\arctan
   \sqrt{\frac{M-2}{2}}\right) \nonumber \\
	& \cdot S\left(2,\frac{\pi}{2}\right) \cdot W\left(1,2,\frac{\pi
   }{4}\right)
	\label{decomposition8}
\end{align}
\end{widetext}
Expression~\eqref{decomposition8} can be easily checked by multiplying it by $Z$ and obtaining the identity matrix.
\subsection{\label{Zdecomposition} Decomposition of $Z$ for $M=8$}

In this section we describe more in detail how the decomposition of $Z^{\dagger}$ can be obtained. We will consider the case of $M=8$, which is reported in Eq.~\eqref{decomposition8}.

We start from the corresponding matrix $Z$, whose expression is reported in Eq.~\eqref{matrixZ8}. This matrix is unitary, and therefore can be decomposed as a sequence of GR \cite{HornJohnson}. To find this decomposition, a handy procedure is to left-multiply $Z$ by $W,\ S$ until we obtain the identity matrix. In short, we should multiply $Z$ by GR that nullify the off-diagonal entries. The sequence of $W,\ S$ then corresponds to the decomposition we are looking for.

To simplify the procedure, we first multiply $Z$ by $W_0 = W(1,\ 2,\ \pi/4)$ and $S_0 = S(2,\ \pi/2)$ to convert the complex entries in the first two rows of $Z$ into their corresponding real and imaginary parts. The matrix becomes
\beq
S_0 \cdot W_0\cdot Z =
\begin{bmatrix}
\frac{\sqrt{2}}{\sqrt{M}} & 0 & \frac{\sqrt{2}\cos(\pi/M)}{\sqrt{M}} &\dots\\
0 & \frac{\sqrt{2}}{\sqrt{M}} & \frac{\sqrt{2}\sin(\pi/M)}{\sqrt{M}} &\\
\sqrt{\frac{M-2}{M}} &  0 & -\frac{2\cos(\pi/M)}{\sqrt{(M-2)M}} & \\
0 & \sqrt{\frac{M-2}{M}} & -\frac{2\sin(\pi/M)}{\sqrt{(M-2)M}} & \\
0 & 0 &  \sqrt{\frac{M-4}{M-2}} \\
\vdots & & & \ddots
\end{bmatrix} .
\label{SWZ}
\eeq
We can then focus on nulling the entry below the diagonal. We need a GR for each of the entries in the first and second columns, $W_1 = W(1,3,\omega_{13})$ and $W_2=W(2,4, \omega_{24})$ respectively, with $\omega_{13} = \omega_{24} =\arctan (\sqrt{(M-2)/2})$. We then obtain 
\beq
{W_2 \cdot W_1 \cdot S_0 \cdot W_0 \cdot Z} =\begin{bmatrix}
1 & 0 & 0 & 0 & \dots \\
0 & 1 & 0 & 0 & \\
0 & 0 & \frac{\sqrt{2}\cos(\frac{\pi}{M})}{\sqrt{M-2}} & -\frac{\sqrt{2}\sin(\frac{\pi}{M})}{\sqrt{M-2}} & \\
0 & 0 & \frac{\sqrt{2}\sin(\frac{\pi}{M})}{\sqrt{M-2}} & \frac{\sqrt{2}\cos(\frac{\pi}{M})}{\sqrt{M-2}} & \\
0 & 0 & \sqrt{\frac{M-4}{M-2}} & 0 & \\
0 & 0 & 0 & \sqrt{\frac{M-4}{M-2}} & \\
\vdots & & & & \ddots
\end{bmatrix} .
\eeq
Note that at this point all the off-diagonal entries in the first and second rows, as well as those in the first and second columns are zero.

If we then multiply the matrix by $W_3 = W(3,4,\pi + \pi/M )$ to nullify the first off-diagonal entry in the fourth row, we obtain a matrix that resembles \eqref{SWZ} except for $M-2$ in place of $M$, i.e. 
\beq
W_3 \cdot W_2 \cdot W_1 \cdot S_0 \cdot W_0 \cdot Z =\begin{bmatrix}
1 & 0 & 0 & 0 & \dots \\
0 & 1 & 0 & 0 & \\
0 & 0 & \frac{\sqrt{2}}{\sqrt{M-2}} & 0 & \\
0 & 0 & 0 & \frac{\sqrt{2}}{\sqrt{M-2}} & \\
0 & 0 & \sqrt{\frac{M-4}{M-2}} & 0 & \\
0 & 0 & 0 & \sqrt{\frac{M-4}{M-2}} & \\
\vdots & & & & \ddots
\end{bmatrix}.
\eeq
The multiplication by $W_3 \cdot W_2 \cdot W_1$ has effectively nullified the left off-diagonal entries in the second and third rows of $S_0 \cdot W_0\cdot Z$.
From here on, we can find triplets of GR $\{W_6,\ W_5,\ W_4\}$ that acts like $\{W_3,\ W_2,\ W_1\}$ to nullify the left off-diagonal entries in rows $5,\ 6$. This procedure can be repeated for the remaining rows and gives the pattern of GR triplets in the decomposition \eqref{decomposition8}.

Once we obtain the final identity matrix, the product of the GR employed $W_9 \cdot W_8 \dots W_3 \cdot W_2 \cdot W_1 \cdot S_0 \cdot W_0$ is a decomposition of $Z^{\dagger}$. The decomposition works for arbitrarily high values of $M$ since the matrix $Z$ has the same structure.

\subsection{Physical realization of Givens Rotations}
\label{physicalRealization}
Without loss of generality, we can consider the measurement of  $\rho_S = \pure{\varphi^{S}}$, with $\ket{\varphi^{S}}$ being a single-photon state,  
\beq
\ket{\varphi^{S}} =  \frac{\ket{0}_L + \ee^{i \varphi} \ket{1}_L}{\sqrt{2}}  = \frac{a^{\dagger(1)}_H + \ee^{i \varphi} a^{\dagger(1)}_V}{\sqrt{2}} \ket{0}, 
\label{singlePhoton}
\eeq
and $\varphi \in [0, 2\pi)$ an unknown phase  to be estimated. Again, $a^{\dagger(1)}_H$ and $a^{\dagger(1)}_V$ are the creator operators for the first path mode, for the horizontal and vertical mode respectively. In the case of a mixed state, the result of the phase measurement follows by linearity from the measurement of the eigenvectors of $\rho_S$.

To define the vector representation of the state in the enlarged Hilbert space $\mathcal{H}$, we collect the coefficients of the creation operators $a^{\dagger(m)}_H,\ a^{\dagger(m)}_V$ and stack them in order in a column. For instance, the input quantum state \eqref{singlePhoton} is represented as 
\beq
\ket{\varphi^{AS}} = 
\ket{e_1^{A}} \otimes \ket{\varphi^{S}} 
\quad
\longrightarrow
\quad
\begin{bmatrix}
\frac{1}{\sqrt{2}} \\
\frac{\ee^{i \varphi}}{\sqrt{2}} \\
0 \\
\vdots \\
0
\end{bmatrix}
\eeq
because the coefficients of $a^{\dagger(1)}_H,\ a^{\dagger(1)}_V$ are placed in the first two items in the column representation, while the zeros are the coefficients of $a^{\dagger(m)}_H,\ a^{\dagger(m)}_V,\ m>1$. 

This representation is useful because in the Hilbert space spanned by the polarizations of a single photon on multiple modes, the mixing of $n$ optical modes is represented by a $n \times n$ unitary matrix \footnote{On the contrary, linear mixing between annihilation operators and creation operators require nonlinear optical interactions, as it results from squeezing transformations.}. As a consequence, the states that define the canonical basis in this representation and in the unitary $Z^{\dagger}$ are single-photon states of some polarization and path modes. The auxiliary state $\ket{e_1^{A}}$ is just the tensor product of many vacuum states corresponding to multiple modes.

In this Hilbert space the converse also holds, i.e., any unitary transformation can be achieved with a set of passive devices such as beam splitters, polarizing beam splitters, waveplates and mirrors \cite{Reck1994}. We will leverage this result to provide a possible realization for the unitaries $S(u, \phi)$ and $W(u,v, \omega)$. 

The unitary $S(u, \phi)$ can be realized with a waveplate of the appropriate thickness where the fast axis is aligned with the horizontal mode and the slow axis with the vertical mode. In this way, the vertical mode gains a phase shift equal to $-\phi$ with respect to the horizontal one. The corresponding transformation given by the waveplate can be expressed as
\beq
\begin{bmatrix}
\hat{a}_H^{\dagger(m)} \\
\hat{a}_V^{\dagger(m)} \\
\end{bmatrix}_{(out)}
=
\begin{bmatrix}
1 & 0 \\
0 & \ee^{-i\phi} \\
\end{bmatrix}
\begin{bmatrix}
\hat{a}_H^{\dagger(m)}\\
\hat{a}_V^{\dagger(m)} \\
\end{bmatrix}_{(in)} \ ,
\label{phaseShiftUnitary}
\eeq
where the index $u$ in $S(u, \phi)$ specifies the column and row associated with $\hat{a}_V^{\dagger(m)}$.

The transformation $W(u,v, \omega)$ can be realized differently depending on whether the modes involved refer to different polarizations of the same rail, or two spatial modes on different  rails. In the first case, a simple rotation equal to $\omega$ of the coordinate system on the polarization plane realizes the transformation, i.e.,
\beq
\begin{bmatrix}
\hat{a}^{\dagger(m)}_H \\
\hat{a}^{\dagger(m)}_V \\
\end{bmatrix}_{(out)}
=
\begin{bmatrix}
\cos (\omega) & \sin(\omega) \\
-\sin (\omega) & \cos(\omega) 
\end{bmatrix}
\begin{bmatrix}
\hat{a}^{\dagger(m)}_H \\
\hat{a}^{\dagger(m)}_V \\
\end{bmatrix}_{(in)} \ .
\label{waveplateMixing}
\eeq
In this case, the indices $u,v$ specify the columns and rows of 
$\hat{a}^{\dagger(m)}_H, \hat{a}^{\dagger(m)}_V $ respectively.
If a mixing between two different spatial modes is required, a beam splitter (BS) with the appropriate transmissivity and reflectivity may be used. The transmissivity and reflectivity may even depend on the polarization, and in this case a partially polarizing beam splitter (PPBS) is required to realize the transformation
\beq
\begin{bmatrix}
\cos (\omega_H) & 0 & \sin(\omega_H) & 0\\
 0 &  \cos (\omega_V) & 0 & \sin(\omega_V) \\
-\sin (\omega_H) &  0 & \cos(\omega_H)  & 0 \\
 0 & -\sin (\omega_V) &  0 & \cos(\omega_V) \\
\end{bmatrix}
\label{ppbsMixing}
\eeq
from the input to the output creation operators enlisted in the column $ [\ \hat{a}^{\dagger(m)}_H,\ \hat{a}^{\dagger(m)}_V,\ \hat{a}^{\dagger(n)}_H, \ \hat{a}^{\dagger(n)}_V  ]^T$. In this case, we are actually implementing the transformation $W(u,v,\omega_H) \cdot W(u',v',\omega_V)$, where $u,v$ point at the coefficients of the horizontal polarizations and $u',v'$ at those of the vertical ones. When $\omega_H = \omega_V$, we recover the transformation of a BS.
An extreme example of PPBS is the polarizing beam splitter (PBS), which completely transmits the horizontal polarizations and reflects the vertical polarizations, and corresponds to the transformation \eqref{ppbsMixing} with $\omega_H = 0,\ \omega_V = \pi/2$.

In order to obtain the implementation of \eqref{decomposition8}, we follow the sequence of the unitary transformations $S(u, \phi),\ W(u,v, \omega)$ from right to left and compose in a cascade the corresponding implementations. Two possible implementations arise, a \emph{direct} one and a \emph{folded} one, which are the topic of the next sections.
\subsection{Direct scheme}
\label{direct}

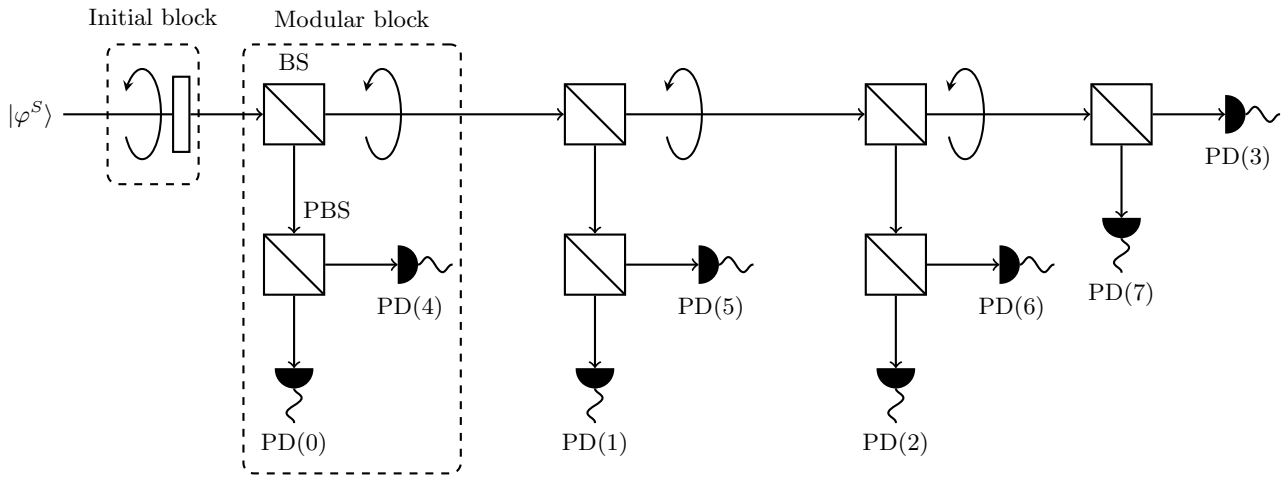
\begin{figure*}  
\centering
	\begin{tikzpicture}[thick]
		\node (o) {$\ket{\varphi^S}$};
		
		\node[draw=black, rectangle, minimum height=10mm, right of=o, node distance=2cm] (l4) {};
		\draw[black] (o) -- node [near end] (rot0) {\AxisRotator} (l4) ;
		\node (initial) [draw=black, dashed, rounded corners, fit =(rot0)(l4)] {};
		\node[above of=initial, yshift=0.3cm]{Initial block};

		\node[pbs, draw=black, rectangle, right of=l4, node distance=1.5cm, minimum size=8mm] (PPBS0) {};
		\node[above of=PPBS0, yshift=-0.3cm] {BS};
		\draw[->,black] (l4) -- (PPBS0);
		
		\node[pbs, draw=black, rectangle, right of=PPBS0, minimum size=8mm, node distance=4cm] (PPBS1) {};
		\draw[->,black] (PPBS0) -- node [pos=0.25] (rot) {\AxisRotator} (PPBS1);
		
		\node[pbs, draw=black, rectangle, right of=PPBS1, minimum size=8mm, node distance=4cm] (PPBS2) {};
		\draw[->,black] (PPBS1) -- node [pos=0.25] {\AxisRotator} (PPBS2);
	
		\node[rectangle, pbs, draw=black, below of=PPBS0, minimum size=8mm, node distance=2cm] (PBS0) {};
		\draw[->,black] (PPBS0) -- (PBS0);
		\node[above of=PBS0, anchor=south west, yshift=-0.5cm] {PBS};
		\node[rectangle, pbs, draw=black, below of=PPBS1, minimum size=8mm, node distance=2cm] (PBS1) {};
		\draw[->,black] (PPBS1) -- (PBS1);
		\node[rectangle, pbs, draw=black, below of=PPBS2, minimum size=8mm, node distance=2cm] (PBS2) {};
		\draw[->,black] (PPBS2) -- (PBS2);
		\node[rectangle, pbs, draw=black, right of=PPBS2, minimum size=8mm, node distance=3cm] (PBS3) {};
		\draw[->,black] (PPBS2) -- node [pos=0.25] {\AxisRotator} (PBS3);

		\node[below of=PBS0, node distance=1.5cm, semicircle, draw=black, fill=black, rotate=180] (pd0) {};
		\draw[->] (PBS0) -- (pd0); 
		\draw[snake=snake,segment amplitude=1mm,segment length=4mm,line after snake=0mm] (pd0.north) -- node (halfsnake1) {} +(0,-.46) node[below, anchor=north] (pd00) {PD(0)} ;
		\node[right of=PBS0, node distance=1.5cm, semicircle, draw=black, fill=black, rotate=-90] (pd1) {};
		\draw[->] (PBS0) -- (pd1); 
		\draw[snake=snake,segment amplitude=1mm,segment length=4mm,line after snake=0mm] (pd1.north) --  +(.46,0) node [below, anchor=north east, yshift=-0.3cm] (pd11) {PD(4)};	
		
		\node (module) [draw=black, dashed, rounded corners, fit =(PPBS0)(rot)(pd00)(pd11)] {};
		\node[above of=module, yshift=2.2cm]{Modular block};			
		
		\node[below of=PBS1, node distance=1.5cm, semicircle, draw=black, fill=black, rotate=180] (pd2) {};
		\draw[->] (PBS1) -- (pd2); 
		\draw[snake=snake,segment amplitude=1mm,segment length=4mm,line after snake=0mm] (pd2.north) -- node (halfsnake1) {} +(0,-.46) node[below, anchor=north] {PD(1)};
		\node[right of=PBS1, node distance=1.5cm, semicircle, draw=black, fill=black, rotate=-90] (pd3) {};
		\draw[->] (PBS1) -- (pd3); 
		\draw[snake=snake,segment amplitude=1mm,segment length=4mm,line after snake=0mm] (pd3.north) --  +(.46,0) node [below, anchor=north east, yshift=-0.3cm] {PD(5)};	
		
		\node[below of=PBS2, node distance=1.5cm, semicircle, draw=black, fill=black, rotate=180] (pd4) {};
		\draw[->] (PBS2) -- (pd4); 
		\draw[snake=snake,segment amplitude=1mm,segment length=4mm,line after snake=0mm] (pd4.north) -- node (halfsnake1) {} +(0,-.46) node[below, anchor=north] {PD(2)};
		\node[right of=PBS2, node distance=1.5cm, semicircle, draw=black, fill=black, rotate=-90] (pd5) {};
		\draw[->] (PBS2) -- (pd5); 
		\draw[snake=snake,segment amplitude=1mm,segment length=4mm,line after snake=0mm] (pd5.north) --  +(.46,0) node [below, anchor=north east, yshift=-0.3cm] {PD(6)};	
		
		\node[below of=PBS3, node distance=1.5cm, semicircle, draw=black, fill=black, rotate=180] (pd6) {};
		\draw[->] (PBS3) -- (pd6); 
		\draw[snake=snake,segment amplitude=1mm,segment length=4mm,line after snake=0mm] (pd6.north) -- node (halfsnake1) {} +(0,-.46) node[below, anchor=north] {PD(7)};
		\node[right of=PBS3, node distance=1.5cm, semicircle, draw=black, fill=black, rotate=-90] (pd7) {};
		\draw[->] (PBS3) -- (pd7); 
		\draw[snake=snake,segment amplitude=1mm,segment length=4mm,line after snake=0mm] (pd7.north) --  +(.46,0) node [text width=2cm, below, yshift=-0.3cm] {PD(3)};	
		
	\end{tikzpicture}
	\caption{Direct scheme for M=8. The qubit $\ket{\varphi^S}$ to be measured enters the optical network from the left and traverses the optical network to give the correct outcome probability. The initial block and a modular block, which is then repeated three times, are enclosed within dashed lines. Each modular block is composed by a BS, a PBS, two photon counters and a polarization plane rotation (depicted with a curved arrow). Each block has two ingoing modes coming from the previous block and two ingoing auxiliary modes in the vacuum state. Two outgoing modes of the BS are directed to the PBS and then to photon counters, while the other two outgoing modes are directed towards the polarization plane rotation and then to the next block. A click in the photon counter PD$(k)$ corresponds to the projection on $P_k=Z^{\dagger} \pure{e_k^{\mathcal{H}}}Z$ in the extended Hilbert space. 
}
	\label{directScheme}
\end{figure*}

The \emph{direct} implementation is designed following the sequence of GR in Eq. \eqref{decomposition8}. Figure \ref{directScheme} depicts the scheme for $M=8$, where the optical network and the photon counters can be clearly recognized.

The scheme presents an \emph{initial block} implementing the unitaries $W\left(1,2,\pi/4 \right)$ and $S\left(2,\pi/2 \right)$. The qubit to be measured is defined on the Cartesian coordinate system of the polarization plane, which is rotated by $\pi/4$ in order to implement $W(1,2,\pi/4)$. In Fig.~\ref{directScheme}, such a rotation is indicated with a curved arrow.

The optical modes $H^{(1)},\ V^{(1)}$ must then go through a quarter-wave plate which realizes $S(2,\pi/2)$. The waveplate, indicated in   Fig.~\ref{directScheme} with a slim rectangular box, is aligned with the new coordinate system, and the same holds for the following components.

The decomposition \eqref{decomposition8} highlights a structure for the matrices following the initial block. In particular, the GR can be grouped in triplets which work on the same group of modes, e.g.
\begin{align}
& W\left(3,4,\omega_{34} \right) \cdot W\left(2,4,\omega_{24} \right) \cdot W\left(1,3,\omega_{13} \right) \ .
	\label{IBmatrices}
\end{align}
The same holds if we add $2k,\ k=0, \ldots, \frac{M}{2}-2$ to the indices of the modes, and employing the angles $\omega_{34}~=~\pi + \frac{\pi}{M}$ and $\omega_{24} = \omega_{13} = \arctan \sqrt{(M-2-2k)/2}$.

This observation suggests a modular implementation of the triplet, which is repeated several times. The modular block is shown within the dashed line in Fig. \ref{directScheme}. 

In general, the unitaries $W\left(2,4,\omega_{24} \right) \cdot W\left(1,3,\omega_{13} \right)$ may be implemented with a PPBS realizing \eqref{ppbsMixing} with $\omega_H = \omega_{13},\ \omega_V = \omega_{24}$, where the horizontal creation operators have indices $1,3$ and the vertical ones have indices $2,4$. However, since $\omega_H = \omega_V = \omega_{24} = \arctan \sqrt{(M-2-2k)/2}$, a BS suffices to implement the transformation. After this, two of the outgoing modes go to a PBS to be separated into horizontal and vertical polarization modes, and then on to photon counters to record a possible click. The other two outgoing modes are mixed with a rotation of the polarization plane, realizing the unitary $W(3,4,\omega_{34} )$ as in \eqref{waveplateMixing}, with $\omega = \omega_{34} = \pi + \frac{\pi}{M}$.

Note that the actual transformation implemented by the BS followed by the rotation of the polarization plane would be (supposing $k=1$)

\beq
\resizebox{\columnwidth}{!}{
$
\begin{bmatrix}
\sqrt{\frac{2}{M}} & 0 & \sqrt{\frac{M-2}{M}} & 0\\
 0 & \sqrt{\frac{2}{M}} & 0 &  \sqrt{\frac{M-2}{M}} \\
\sqrt{\frac{M-2}{M}} \cos(\frac{\pi}{M}) & \sqrt{\frac{M-2}{M}} \sin(\frac{\pi}{M})  & -\sqrt{\frac{2}{M}} \cos(\frac{\pi}{M})   & -\sqrt{\frac{2}{M}} \sin(\frac{\pi}{M})  \\
-\sqrt{\frac{M-2}{M}} \sin(\frac{\pi}{M}) & \sqrt{\frac{M-2}{M}} \cos(\frac{\pi}{M}) & \sqrt{\frac{2}{M}} \sin(\frac{\pi}{M}) & -\sqrt{\frac{2}{M}} \cos(\frac{\pi}{M}) 
\end{bmatrix}.
$
}
\eeq

However, two of the input modes of the PPBS are vacuum states, and the effective transformation from the coefficients of $[\hat{a}^{\dagger(m)}_H,\ \hat{a}^{\dagger(m)}_V ]^T$ to those of $[\ \hat{a}^{\dagger(m)}_H,\ \hat{a}^{\dagger(m)}_V,\ \hat{a}^{\dagger(n)}_H,\ \hat{a}^{\dagger(n)}_V ]^T$ 
results
\beq
\resizebox{\columnwidth}{!}{
$
\begin{bmatrix}
\hat{a}^{\dagger(m)}_H \\
\hat{a}^{\dagger(m)}_V \\
\hat{a}^{\dagger(n)}_H \\
\hat{a}^{\dagger(n)}_V 
\end{bmatrix} = 
\begin{bmatrix}
\sqrt{\frac{2}{M}} & 0 \\
 0  & \sqrt{\frac{2}{M}} \\
\sqrt{\frac{M-2}{M}} \cos(\frac{\pi}{M}) & \sqrt{\frac{M-2}{M}} \sin(\frac{\pi}{M}) \\
-\sqrt{\frac{M-2}{M}} \sin(\frac{\pi}{M}) & \sqrt{\frac{M-2}{M}} \cos(\frac{\pi}{M}) 
\end{bmatrix}
\begin{bmatrix}
\hat{a}^{\dagger(m)}_H \\
\hat{a}^{\dagger(m)}_V
\end{bmatrix}.
$
}
\label{equivalentTransformation}
\eeq

The horizontal and vertical polarizations of mode $m$ then goes to a PBS followed by two photon counters, while those in mode $n$ go to the next modular block, or to additional photon counters in the case of the last module. 

As a side note, we would like to point out that, in general, photonic implementations with BS require the rails to be swapped and brought close in order to perform the unitary operation on adjacent modes. The direct scheme, as well as the folded scheme, are free of this issue as can be seen from the schematics of Figs.~\ref{directScheme} and \ref{foldedScheme}. This is also true for arbitrarily high $M$ since this property originates from the pattern of GR triplets in the decomposition of $Z^{\dagger}$ and their modular implementation with a BS and a polarization plane rotation.

\subsection{Folded scheme}
\label{folded}

The folded scheme is a variation of the direct scheme which comes from the following considerations. 

Firstly, an experimental implementation of the direct scheme requires a number of photon counters equal to the number of outcomes $M$, which may become highly expensive to realize if a fine resolution of the phase is required, i.e., a large number $M$. It is therefore worthwhile exploring the possibility of reducing the number of devices required.

Secondly, the photon counters and the modular blocks in general are not ``used'' at the same time, but at different times as the photon will click later in PD$(k)$ with respect to PD$(0)$. This opens up the possibility to exploit ``recursive'' schemes that reuse the same modular block in different time slots.

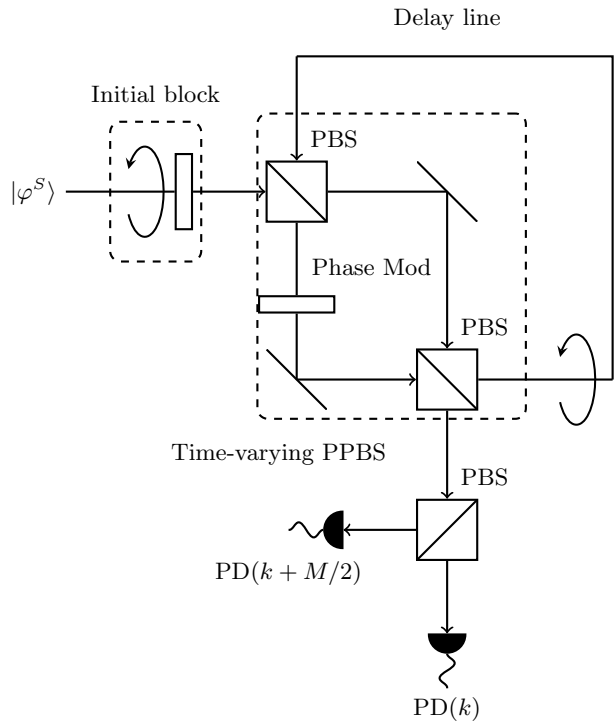
\begin{figure*}[htb]
\centering
	\begin{tikzpicture}[thick]
		\node (o) {$\ket{\varphi^S}$};
		
		\node[draw=black, rectangle, minimum height=10mm, right of=o, node distance=2cm] (l4) {};
		\draw[black] (o) -- node [near end] (rot0) {\AxisRotator} (l4) ;
		\node (initial) [draw=black, dashed, rounded corners, fit =(rot0)(l4)] {};
		\node[above of=initial, yshift=0.3cm]{Initial block};

		\node[pbs, draw=black, rectangle, right of=l4, node distance=1.5cm, minimum size=8mm] (PBS0) {};
		\node[above of=PPBS0, yshift=-0.3cm, xshift=0.5cm] (pbsLabel0) {PBS};
		\draw[->,black] (l4) -- (PBS0);
		
		\node[pbs, right of=PBS0, node distance=2cm, minimum size=8mm] (mirror0) {};
		\draw[black] (PBS0) -- (mirror0.center);
		\node[draw=black, rectangle, below of=PBS0, node distance=1.5cm, minimum width=10mm] (phaseMod) {};
		\node[above of=phaseMod, yshift=-0.5cm, xshift=1cm] {Phase Mod};
		\draw[black] (PBS0) -- (phaseMod);
		
		\node[pbs, below of=phaseMod, node distance=1cm, minimum size=8mm] (mirror1) {};
		\draw[black] (phaseMod) -- (mirror1.center);
		\node[pbs, draw=black, rectangle, right of=mirror1, node distance=2cm, minimum size=8mm] (PBS1) {};
		\node[above of=PBS1, yshift=-0.3cm, xshift=0.5cm] (pbsLabel1) {PBS};
		\draw[->,black] (mirror1.center) -- (PBS1);
		\draw[->,black] (mirror0.center) -- (PBS1);
		
		\node[rectangle, antiPbs, draw=black, below of=PBS1, minimum size=8mm, node distance=2cm] (PBS2) {};
		\draw[->,black] (PBS1) -- (PBS2);
		\node[above of=PBS2, yshift=-0.3cm, xshift=0.5cm] {PBS};

		\node[below of=PBS2, node distance=1.5cm, semicircle, draw=black, fill=black, rotate=180] (pd0) {};
		\draw[->] (PBS2) -- (pd0); 
		\draw[snake=snake,segment amplitude=1mm,segment length=4mm,line after snake=0mm] (pd0.north) -- node (halfsnake1) {} +(0,-.46) node[below, anchor=north] (pd00) {PD$(k)$} ;
		\node[left of=PBS2, node distance=1.5cm, semicircle, draw=black, fill=black, rotate=90] (pd1) {};
		\draw[->] (PBS2) -- (pd1); 
		\draw[snake=snake,segment amplitude=1mm,segment length=4mm,line after snake=0mm] (pd1.north) --  +(-.46,0) node [below, anchor=north, yshift=-0.3cm] (pd11) {PD$(k+M/2)$};	
		
		\node (module) [draw=black, dashed, rounded corners, fit =(PBS0)(PBS1)(pbsLabel0)(pbsLabel1)] {};
		\node[left of=module, xshift=-0.5cm, yshift=-2.5cm]{Time-varying PPBS};			
	
		\node[right of=PBS1, node distance=2.2cm] (out1) {};
		\node[above of=PBS0, node distance=1.8cm] (out0) {};
		\draw[black] (PBS1) -- node [near end] {\AxisRotator} (out1.center);
		\draw[->, black] (out0.center) -- (PBS0);
		\draw[black] (out1.center) |- (out0.center);
		\node[above of=out0, yshift=-0.5cm, xshift=2cm] {Delay line};
		
	\end{tikzpicture}
	\caption{Folded scheme for the phase measurement. The qubit $\ket{\varphi^S}$ to be measured travels through an initial block and enters an optical loop defined by the interferometer and the delay line. The initial block and the interferometer, which implements a time-varying BS, are enclosed within dashed lines. The optical loop, composed by the interferometer, PBS, photon counters and the polarization-plane rotation (depicted with a curved arrow) corresponds to the modular block of the direct scheme.	The photon exits the loop via the second BS in the interferometer, and its polarizations are splitted with a PBS and directed to photon counters. The measurement outcome $k$ [or $k+M/2$] is obtained when a click in PD$(k)$ [PD$(k+M/2)$] is recorded in the $(k+1)$-th time slots, which is defined as the time interval that takes the photon to travel in the loop.
	}
	\label{foldedScheme}
\end{figure*}

The folded scheme is obtained by putting a delay line after the first modular block, connecting the output modes of this block to the input modes of the PPBS, effectively ``folding'' all the blocks onto the first one. Fig.~\ref{foldedScheme} shows a possible implementation of the scheme.

Note that since the parameters $\omega_H,\ \omega_V$ change from one modular block to another, a time-varying BS is required, which may be realized with an interferometer  with the appropriate phase shift on one arm. This interferometer is singled out inside the dashed enclosure in Fig.~\ref{foldedScheme}. 

The photon to be measured enters the loop and in the following $M/2$ time slots (each lasting a round trip time in the delay line) it exits towards the PBS and the photon counters. The outcome depends both on the polarization and the time slot, and it corresponds to $k$ and $k+M/2$ for a click recorded in the $(k+1)$-th time slot in the photon counter associated respectively with the horizontal and vertical polarization. 

A comment on the experimental feasibility of these schemes is in order. 
As a matter of fact, the direct scheme can straightforwardly
be realised with bulk optics or in integrated optical 
circuits. On the other hand, the folded scheme requires a careful 
design of the delay line. In particular, its length defines the 
time slot where a photon can be recorded at the photon counters, 
and must at least amount to the temporal span of the single photon 
plus the time required to the time-varying BS to adjust its
parameters. This latter duration is the slower constraint, with 
commercial devices that reports switching frequencies of the order 
of tens of MHz in their datasheet. The resulting length for the delay 
line is of the order of tens of meters, which are feasible to realize 
in a lab. The dead time of photon counter, which may blind successive 
time slots, does not impact the current measurement, thought it may 
affect the time slots in the following one. This can be solved by imposing 
an idle time interval between consecutive measurements, which reduces the overall 
rate of measurements that can performed. However, for a proof-of-principle 
experimental test this is usually not an issue.

\section{Conclusions}
In conclusion, we have addressed the optical implementation of
POVM corresponding to the optimal $M$-outcomes discrimination 
of the polarization state of a single photon.  In particular,
we have found an explicit Naimark extension and optical
implementation for any $M=2^N,\ N > 1$, so that the resolution 
of the estimated phase can be arbitrarily small. 
\par
The measurement scheme has been devised to estimate the phase of the 
polarization of a single photon. The single photon passes through an 
optical network towards a set of photon counters, providing information
about which path was taken, depending on its polarization.
The optical network is defined by the unitary obtained from the projectors, 
and it is realized as a sequence of modular blocks that reflects the structure 
of the unitary decomposition in GR. Each block is a combination of beam 
splitters and waveplates that act on multiple polarization modes. The 
photon counters are placed at the outgoing modes and at each recorded click 
they assign the corresponding outcome. 
\par
We have provided the analytical expression for both the Naimark extension 
and its decomposition in GR, and we have proposed an implementation for
the measurement scheme of the polarization, but other phase measurements 
can in principle be realized. Our results pave the way for realistic
implementations of the canonical phase POVM for single-photon states 
and for extension to high-dimensional Hilbert spaces.

\begin{acknowledgments}
N. Dalla Pozza thanks G. Vallone, M. Avesani and J. Tinsley for useful discussions 
and comments. M. G. A. Paris is member of GNFM-INdAM and thanks 
S. Olivares and S. Cialdi for discussions.
\end{acknowledgments}


\pagebreak
\bibliography{ndppp}


\end{document}